\pdfoutput=1
\documentclass[12pt,letterpaper,floatfix]{article}
\usepackage{amsmath,amssymb,bbm,color}
\usepackage{graphicx}
\usepackage{epsfig}
\usepackage{euscript}
\usepackage{pslatex}
\usepackage{fancybox}
\usepackage{enumerate}
\usepackage{widetext}
\usepackage{subfigure}
\let\sstl=\scriptscriptstyle
\newcommand{\Meu}{\EuScript{M}}

\newcommand{\KK}{${\cal KK}$}

\def\st{\hbox{}} %% hbox is lower
\usepackage[figuresright]{rotating}

\begin{document}
\begin{titlepage}
%\begin{frontmatter}

%% Title, authors and addresses

%% use the tnoteref command within \title for footnotes;
%% use the tnotetext command for the associated footnote;
%% use the fnref command within \author or \address for footnotes;
%% use the fntext command for the associated footnote;
%% use the corref command within \author for corresponding author footnotes;
%% use the cortext command for the associated footnote;
%% use the ead command for the email address,
%% and the form \ead[url] for the home page:
%%
%% \title{Title\tnoteref{label1}}
%% \tnotetext[label1]{}
%% \author{Name\corref{cor1}\fnref{label2}}
%% \ead{email address}
%% \ead[url]{home page}
%% \fntext[label2]{}
%% \cortext[cor1]{}
%% \address{Address\fnref{label3}}
%% \fntext[label3]{}
\begin{center}
%\dochead{}
\begin{flushleft}
{\small\bf BU-HEPP-17-01, IFJPAN-IV-2017-15, May, 2017}
\end{flushleft}
\vspace{18mm}
%% Use \dochead if there is an article header, e.g. \dochead{Short communication}

{\bf\Large Systematic Studies of Exact ${\cal O}(\alpha^2L)$ CEEX EW Corrections in a Hadronic MC for Precision $Z/\gamma^*$ Physics at LHC Energies}\\
\vspace{2mm}
%% use optional labels to link authors explicitly to addresses:
%% \author[label1,label2]{<author name>}
%% \address[label1]{<address>}
%% \address[label2]{<address>}

{S. Jadach$^a$, ~B.F.L. Ward$^b$,~Z. W\c{a}s$^a$,~S.A. Yost$^c$}\\
%\author{B.F.L. Ward$^b$}
%\author[label2]{S. Jadach}
%\author{Z. Was$^a$}
%\Newlabel{label1}{{Institute of Nuclear Physics, Cracow, PL}
%\begin{center}
{$^a$Institute of Nuclear Physics Polish Academy of Sciences, Cracow, PL}\\
{$^b$Baylor University, Waco, TX, USA}\\
{$^c$The Citadel, Charleston, SC, USA}\\
%\address[label2]{Institute of Nuclear Physics, Cracow, PL}
\end{center}
\centerline{\bf Abstract}
%% Text of abstract
With an eye toward the precision physics of the LHC, such as the recent measurement of $M_W$ by the ATLAS Collaboration, we present here systematic studies relevant to the assessment of the expected size of multiple photon radiative effects in heavy gauge boson production with decay to charged lepton pairs. We use the new version 4.22 of {\KK}MC-hh so that we have coherent exclusive exponentiation (CEEX) electroweak (EW) exact ${\cal O}(\alpha^2 L)$ corrections in a hadronic MC and control
over the corresponding EW initial-final interference (IFI) effects  as well. In this way, we illustrate the interplay between cuts of the type used in the measurement of $M_W$ at the LHC and the sizes of the expected responses of the attendant higher order corrections. We find that there are per cent to per mille level effects in the initial-state radiation, fractional per mille level effects in the IFI and per mille level effects in the over-all ${\cal O}(\alpha^2 L)$ corrections that any treatment of EW corrections at the per mille level should consider. Our results have direct applicability to current LHC experimental data analyses.
%\end{abstract}
%\begin{keyword}
%% keywords here, in the form: keyword \sep keyword
%Coherent Exclusive Exponentiation EW Corrections Precision LHC Physics
%% MSC codes here, in the form: \MSC code \sep code
%% or \MSC[2008] code \sep code (2000 is the default)
%\end{center}
%\end{keyword}

%\end{frontmatter}
\end{titlepage}

%\begin{document}

%%
%% Start line numbering here if you want
%%
% \linenumbers
%%%Bornxsec=886.983 +- 0.040 pb (before cuts), 311.663+-0.043pb after cuts
%% main text>
\section{Introduction}
%\label{intro}
With large data samples already at 7TeV and even larger ones at 8TeV and 13TeV, the LHC experiments are well into the era of precision QCD$\otimes$EW physics for processes
such as single heavy gauge boson production with decay to lepton pairs. As an example, the ATLAS Collaboration has recently used their 7TeV data samples to measure the mass of the
W boson with the result~Ref.~\cite{atlasmw-17}: 
$$m_W     =   80370 \pm 7 (\text{stat.}) \pm 11 (\text{exp. syst.}) \pm 14 (\text{mod. syst.}) \text{MeV}\\
=   80370 \pm 19 \text{MeV},$$
where the first uncertainty is statistical, the second is the experimental systematic uncertainty,  and the third is the physics-modeling systematic uncertainty.
The result itself is already one of the most precise single measurements of $m_W$~\cite{cdf-d0} and bodes well, given the remaining data samples that have yet to be analyzed, for a new level of precision in the observable $m_W$ in LHC physics. The essential feature is high precision in measurements of lepton directions in detectors of hadronic colliders, see eg.ref.~\cite{d0-a}.
\par
The error budget in this pioneering measurement of $m_W$ reveals that the modeling systematic error is the largest one at 14 MeV. As we anticipate the type of error the currently available data will yield when it has all been analyzed, we see that the statistical component, now at 7 MeV, will drop by a factor $\sim 4$ to the 1.8 MeV regime. Thus, it is imperative to reduce the 
large modeling error in kind as much as it is possible. Specifically, in the measurement of $M_W$ by the ATLAS Collaboration~\cite{atlasmw-17}, properties of the W production and decay systematics, such as the momentum resolution scale uncertainty, are estimated by comparing with the analogous systematics for the $Z/\gamma^*$ production and decays. The uncertainty on the corresponding EW corrections then contribute
to these systematics. In what follows, we explore the possible role of the new exact\footnote{Here and henceforth we use the notation $L$ for the big log in the respective radiative effects discussed with $L=\ln(Q^2/{m_f}^2)$ where $Q$ is the hard momentum scale of the radiation and $m_f$ is the rest mass of the light fermion $f$ which emits that radiation.} ${\cal O}(\alpha^2L)$ coherent exclusive exponentiation (CEEX) electroweak (EW) corrections in \KK{MC}-hh~\cite{kkmchh} in this context. We note that
some results into that direction were already obtained with the help of {\KK} MC 4.19~\cite{Jadach:1999vf,Jadach:2000ir}, see eg. Ref.~\cite{tkodoan:2013}.  At that time it was impossible to take simultaneously into account QCD initial state parton shower effects. The papers  
in Refs.~\cite{erw-zw:2016-1} or
~\cite{erw-zw:2016-2} recall and use old observations of spin amplitudes for one and two parton emissions: the electroweak part can be quite well factorized from the pp QCD dominated amplitudes.\par

Continuing in this way, we note that in Ref.~\cite{atlas-1612-03016}, the precision measurements are reported for the Z differential spectra in which in the Z peak region  have 2.5 per mille
statistical errors with NLO EW corrections at the 2-3 per mille level in the central rapidity regime when the final state radiation (FSR) is unfolded from the data using PHOTOS~\cite{Barberio:1990ms,Barberio:1994qi,photos2:1994,Golonka:2005pn,MsCGolonka,Golonka:2006tw,Davidson:2010ew}\footnote{We stress that the PHOTOS next-leading log precision tests in Ref.~\cite{Golonka:2006tw} are those of {\KK} MC 4.19~\cite{Jadach:1999vf,Jadach:2000ir} and that in Ref.~\cite{Davidson:2010ew} pair emission is realized in PHOTOS.}. At this level of precision, it is important to assess all possible higher order EW effects that may enter at the per mille level. Such an assessment is what we do in the following.\par
In fact, in Ref.~\cite{atlas-1612-03016} the NLO EW corrections which are not unfolded with PHOTOS 
are implemented in the so-called additive approach, in which one uses (see eqs. (13) - (15) in Ref.~\cite{atlas-1612-03016}) 
\begin{equation}\sigma^{\text{NLO}\;\text{EW}}_{\text{PO-QCD}} = \sigma^{\text{LO}\;\text{EW}}_{\text{PO-QCD}}\left(1+\frac{K^{\text{EW}}-1}{K_{\text{QCD}}}\right),
\label{eq1}
\end{equation}
to implement the EW NLO corrections as opposed to the so-called multiplicative or factorized approach in which one uses
\begin{equation}\sigma^{\text{NLO}\;\text{EW}}_{\text{PO-QCD}} = \sigma^{\text{LO}\;\text{EW}}_{\text{LO}\;\text{QCD}} K^{\text{EW}} K_{\text{QCD}}
\label{eq2}
\end{equation}
where the EW and QCD K factors are $K^{\text{EW}}=\sigma^{\text{NLO}\;\text{EW}}_{\text{LO}\;\text{QCD}}/\sigma^{\text{LO}\;\text{EW}}_{\text{LO}\;\text{QCD}}$ and $K_{\text{QCD}} = \sigma^{\text{LO}\;\text{EW}}_{\text{PO-QCD}}/$\\
$\sigma^{\text{LO}\;\text{EW}}_{\text{LO}\;\text{QCD}}$
where $\text{PO-QCD}=\text{NNLO}\;\text{QCD}$ in Ref.~\cite{atlas-1612-03016}. Here we use an obvious notation for the various orders of the respective cross sections.
We would point out that the dominant parts of the corrections, their leading log parts, are strongly ordered and thereby independently realized so that they must factorize. This suggests that the multiplicative/factorized approach is more efficient at summing up higher order effects beyond NLO~\cite{den-ditt1211.5078}. In what follows, we will use {\KK}MC-hh to realize exact ${\cal O}(\alpha^2L)$ EW corrections in single $Z/\gamma^*$ production in a hadronic MC in what amounts to a factorized treatment of the EW and QCD corrections. The results in Ref.~\cite{dittmr-1,dittmr-2} suggest that the non-factorizable corrections are small.\par

The need to consider higher order EW corrections beyond NLO can also be seen in the ATLAS results in Ref.~\cite{atlas-8TeV} in which differential spectra for single $Z/\gamma^*$
production with decay to lepton pairs are presented with per mille level statistical errors in the Z peak regime. Such precision asks for the treatment of all EW effects that
enter at the per mille level as we present in what follows. Indeed, in Ref.~\cite{atlas-8TeV} the FSR is unfolded from the data along with detector effects using PHOTOS.
The multiplicative implementation of (\ref{eq2}) is used to introduce NLO EW corrections to the NNLO QCD predictions of DYNNLO~\cite{dynnlo} but the comparison with the data is inconclusive as to whether the NLO EW corrections improve the agreement between theory and experiment. We would note that, with {\KK}MC-hh, one now has the option of unfolding exact ${\cal O}(\alpha^2L)$ CEEX EW
corrections from the data\footnote{Here, we have in mind the analog of the unfolding of FSR from the data using PHOTOS in Ref.~\cite{atlas-8TeV}, for example.}. This would afford a much more complete test of the Standard Model (SM) EW-QCD theory. We encourage experimentalists to make such a test.\par

We also observe that the results in Ref.~\cite{atlas-7TEV} feature 2 per mille level statistical errors on the differential spectra of single $Z/\gamma^*$ production at the LHC at 7TeV.
The unfolding of FSR is done with PHOTOS and cross-checked with SHERPA~\cite{sherpa}. This results in a 0.3\% error assessment across the $p_T$ spectrum for the muon pair case and a 0.1\% error in the electron pair case. At this level of precision, we would suggest the unfolding with the {\KK}MC-hh with exact ${\cal O}(\alpha^2L)$ CEEX EW
corrections would be in order, as we illustrate in what follows.\par

In Ref.~\cite{cms-8TEV} per mille level statistical errors are reported for the $Z/\gamma^*$ $p_T$ spectra in the regime of $p_T < 60\;\text{GeV}$ given with a bin size of 20 GeV. The systematic error from FSR is estimated from the difference between an exact ${\cal O}(\alpha)$ result and a soft-collinear approach and results in a per mille level contribution
to the systematic errors in differential spectra. The use of {\KK}MC-hh to address the other EW effects that enter at this level would therefore be appropriate, as we shall illustrate in the following.\par

Continuing in this way, we observe that, in Ref.~\cite{lhcb-7TEVmumu}, the error due to FSR is estimated by comparing the results from Herwig++~\cite{hwg++} and Pythia8~\cite{py8}
with the result that in differential spectra the FSR uncertainty varies between 0.3\% and 3\%. In Ref.~\cite{lhcb-8TEV}, the same approach is used for the FSR correction and again with the result that 0.3\% FSR errors obtain in differential spectra in some regions of phase space. These are again cases where the effects of other EW corrections that
enter at the per mille level could be significant, as we will illustrate with {\KK}MC-hh in what follows.\par

Note that comparisons of results from two different Monte Carlo programs does not exhaust the topic of systematic errors. For that
purpose comparisons between results of distinct physics assumptions  need to be performed. This needs to be performed after technical and statistical errors are found to be under control. Only for the case of KKMC-hh we will explore the context.\par

We also call attention to the studies  in Refs.~\cite{wack1,wack2,ditt1,fulv1,ditt2} and  in Ref.~\cite{vicini-wack:2016} on the expected sizes of the EW corrections in LHC observables. We address the detailed relationship between our {\KK}MC-hh results and those in these latter references elsewhere~\cite{elswh}. Historically, for neutral current Drell-Yan processes, Herwig~\cite{HERWIG}, Pythia~\cite{py8,Pythia}, Herwig++~\cite{hwg++} and Sherpa~\cite{sherpa} have featured QED radiative effects in the context of parton showers: the leading-log QED shower is available in Herwg, Herwig++, Pythia and Sherpa and final state YFS exponentiated radiation for decays is available in Herwig++ and in Sherpa. Recently~\cite{sherpa2,recola,opnlps}, Sherpa, Recola and OpenLoops authors have made available exact ${\cal O}(\alpha)$  EW corrections and exact NLO QCD corrections to such Drell-Yan processes as an option with parton showers. In the Powheg framework, the corresponding exact ${\cal O}(\alpha)$  EW corrections and exact NLO QCD corrections are available as presented in Ref.~\cite{powheg1}. We note that  SANC~\cite{sanc} features NLO QCD and NLO EW corrections to neutral current Drell-Yan processes and that FEWZ~\cite{fewz} features exact NNLO QCD and
the exact ${\cal O}(\alpha)$  EW corrections to such processes. Finally, we call attention again to the exact ${\cal O}(\alpha\alpha_s)$ non-factorizable corrections to the neutral current Drell-Yan process already referenced in Refs.~\cite{dittmr-1,dittmr-2}, which are available, along with the NLO QCD and NLO EW corrections, in the MC integrator program RADY.\par

One further point requires some discussion. In the structure function approach to EW corrections in hadronic collisions, one is led naturally to the inclusion of QED kernels in the DGLAP-CS~\cite{dglap1,dglap2,dglap3,dglap4,dglap5,dglap6,dglap7,dglap8,dglap9} equations with a photon parton inside the proton at the LHC/FCC. The origin of the photon partons in the proton is radiation by quarks and ant-quarks~\cite{mstw-mass} -- a proton at rest does not contain photons as bound-state constituents. Hence, in our approach, such contributions are calculated as part of the set of processes ${\llap{\phantom q}^{\sstl(}\bar q^{\sstl)}{}}+{\llap{\phantom q}^{\sstl(}\bar q^{\sstl)}{}}\rightarrow {\llap{\phantom q}^{\sstl(}\bar q^{\sstl)}{}}+{\llap{\phantom q}^{\sstl(}\bar q^{\sstl)}{}}+\ell +\bar{\ell}$,$\; q= u,\; d,\; s,\; c,\; b, \;\ell=e^-,\;\mu^-,\;\tau^-$, with the ATLAS cuts
on the lepton pair as given in Ref.~\cite{atlasmw-17}, which we repeat below. With these cuts, these processes are ${\cal O}(\alpha^2)$ in our analysis, as it can be seen already
from the results in Ref.~\cite{ditt2}, where the sum of the photon-induced processes essentially vanishes (i.e., is very small) in the region of interest for the invariant lepton pair mass
distribution between 80 GeV and 100 GeV (See Fig. 12 in Ref.~\cite{ditt2}). We will take up these ${\cal O}(\alpha^2)$ effects elsewhere~\cite{elswh}.\par

The paper is organized as follows. In the next section we give a brief recapitulation of the physics and methodology in the {\KK}MC-hh MC, as they are still not a generally familiar.
In Section 3 we illustrate the effect of the EW corrections in {\KK}MC-hh in the context of the type of acceptance used by ATLAS in their use of single $Z/\gamma^*$ events with decays to lepton pairs in their precision measurement of $m_W$ in Ref.~\cite{atlasmw-17}. %%In Section 4, we make contact with the studies in Ref.~\cite{vicini-wack}. 
In Section 4, we summarize our findings in view of our discussion in this Introduction.\par

\section{Recapitulation of the Physics in {\KK}MC-hh}

{\KK}MC-hh is the union of two developments in the Monte Carlo event generator approach to precision theoretical physics for high energy colliding beam devices: The exact amplitude-based  CEEX/EEX YFS MC approach to EW higher order corrections pioneered in Refs.~\cite{Jadach:1993yv,Jadach:1999vf,Jadach:2000ir,Jadach:2013aha} and the QCD parton shower hadron MC approach pioneered in Refs.~\cite{sjos-sh,HERWIG}. Here, EEX denotes exclusive exponentiation as originally formulated by Yennie, Frautschi and Suura (YFS) in Ref.~\cite{yfs:1961}. In the discussion which follows, we will use the Herwig6.5~\cite{HERWIG} MC for the parton shower realization but we continue to stress that the use of any parton shower MC which accepts LHE~\cite{lhe-format} input is allowed in {\KK}MC-hh studies. To give a brief recapitulation of the physics in {\KK}MC-hh we proceed as follows.\par

We start with the master formula for the CEEX realization of the higher corrections to the SM~\cite{SM1,SM2,SM3,SM4} EW theory. 
For completeness, let us recall that the CEEX realization is amplitude level coherent exclusive exponentiation whereas the EEX realization is exclusive exponentiation at the squared amplitude level. Considering the prototypical process 
$q\bar{q}\rightarrow \ell\bar{\ell}+n\gamma, \; q=u,d,s,c,b,t,\ell=e,\mu,\tau,\nu_e,\nu_\mu,\nu_\tau,$ we have the cross section formula
\begin{equation}
\sigma =\frac{1}{\text{flux}}\sum_{n=0}^{\infty}\int d\text{LIPS}_{n+2}\; \rho_A^{(n)}(\{p\},\{k\}),
\label{eqn-hw2.1-1}
\end{equation}
where $\text{LIPS}_{n+2}$ denotes Lorentz-invariant phase-space for $n+2$ particles and $A=\text{CEEX},\;\text{EEX}$. The incoming and outgoing fermion momenta are abbreviated as $\{p\}$ and the $n$ photon momenta are denoted by $\{k\}$.
Note, that thanks to use of conformal symmetry, full $2+n$ body  phase space is covered without any
approximations. Details of the algorithm are covered in Ref.~\cite{Jadach:1999vf}.
To be specific, we note from Refs.~\cite{Jadach:2000ir,Jadach:1999vf,kkmchh} that 
\begin{equation}
\rho_{\text{CEEX}}^{(n)}(\{p\},\{k\})=\frac{1}{n!}e^{Y(\Omega;\{p\})}\bar{\Theta}(\Omega)\frac{1}{4}\sum_{\text{helicities}\;{\{\lambda\},\{\mu\}}}
\left|\Meu\left(\st^{\{p\}}_{\{\lambda\}}\st^{\{k\}}_{\{\mu\}}\right)\right|^2.
\label{eqn-hw2.1-2}
\end{equation}
The corresponding formula for the $A=\text{EEX}$ case is also given in Refs.~\cite{Jadach:2000ir,Jadach:1999vf}. $Y(\Omega;\{p\})$ is the YFS infrared exponent and the attendant infrared integration limits are specified by the region $\Omega$ and its characteristic function
$\Theta(\Omega,k)$ for a photon of energy $k$, with $\bar\Theta(\Omega;k)=1-\Theta(\Omega,k)$ and $$\bar\Theta(\Omega)=\prod_{i=1}^{n}\bar\Theta(\Omega,k_i).$$
For the definitions of the latter functions as well as the CEEX amplitudes $\{\Meu\}$ we refer the reader to  Refs.~\cite{Jadach:1999vf,Jadach:2000ir,Jadach:2013aha}. 
{\KK}MC-hh inherits from \KK MC 4.22 the exact
${\cal O}(\alpha)$ EW corrections implemented using the  DIZET6.2.1 EW library from the semi-analytical
program ZFITTER~\cite{zfitter1,zfitter6:1999}. As the respective implementation is described in Ref.~\cite{Jadach:2000ir} we do not repeat it here.
We stress that the CEEX amplitudes $\{\Meu\}$ in (\ref{eqn-hw2.1-2}) are exact in ${\cal O}(\alpha^2 L^2, \alpha^2L)$ in {\KK}MC-hh.
\par

The union with the parton shower MC approach is facilitated via the standard Drell-Yan formula for the $pp\rightarrow Z/\gamma^*+X\rightarrow \ell\bar{\ell}+X'$, $\ell = e^-,\mu^-$:
\begin{equation}
\sigma_{\text{DY}}=\int dx_1dx_2\sum_i f_i(x_1)f_{\bar{i}}(x_2)\sigma_{\text{DY},i\bar{i}}(Q^2)\delta(Q^2-x_1x_2s),
\label{eqn-hw2.1-3}
\end{equation}
where the subprocess for the $i$-th $q\bar{q}$ annihilation with $\hat{s}=Q^2$ when the pp cms energy squared is $s$
is given in a conventional notation for parton densities $\{f_j\}$. For a given QCD parton shower MC, {\KK}MC-hh receives multiple gluon radiation via the backward evolution~\cite{sjos-sh} 
for the densities as specified in
(\ref{eqn-hw2.1-3}). This backward evolution then also gives {\KK}MC-hh the corresponding hadronization for that shower.
While we use in what follows the Herwig6.5 shower MC for this phase of the event generation, we continue to stress that, as the Les Houches Accord format is also available for the hard processes generated in {\KK}MC-hh before the shower, all shower MC's which use that format can be used for the shower/hadronization part of the simulation.\par

\section{CEEX Exact ${\cal O}(\alpha^2L)$ EW Effects from {\KK}MC-hh for the ATLAS Acceptance for $Z/\gamma^*$ Decays to Lepton Pairs Used in the Measurement of $M_W$}
As we have noted, in their pioneering measurement of $M_W$, the ATLAS Collaboration~\cite{atlasmw-17} estimates properties of the W production and decay systematics by comparing with the analogous systematics for the $Z/\gamma^*$ production and decays, so that the corresponding EW corrections uncertainty contributes to these systematics. What we will do in this section is to the use the $Z/\gamma^*$ cuts from systematics studies done by ATLAS in their $m_W$ analysis to illustrate the size of the new 
higher order EW effects
in {\KK}MC-hh in the context of those cuts.\footnote{We understand that in the actual ATLAS analysis~\cite{atlasmw-17} for the $Z$ spectra the effects of the $m_Z$ uncertainty and absence of fermion pair radiation in the calibration systematic uncertainties were included but all other EW effects were neglected~\cite{prvte-commun}.} \par

The ATLAS cuts on the $Z/\gamma^*$ production and decay to lepton pairs employed in Ref.~\cite{atlasmw-17} are as follows:
$$ 80\;\text{ GeV}<M_{\ell\ell}<100\; \text{GeV},\; P^{\ell\ell}_{T}< 30\;\text{GeV},\; $$
where both members of the decay lepton pair satisfy $$ P^{\ell}_{T}> 25\;\text{ GeV},\; |\eta_\ell| < 2.4. $$ 
Here, we have defined  $M_{\ell\ell}$ as the lepton pair invariant mass, $P^{\ell\ell}_{T}$ as the transverse momentum of the lepton pair,  $P^{\ell}_{T}$ as the transverse momentum of the lepton or anti-lepton $\ell$, and $\eta_\ell$ as the pseudorapidity of the lepton or anti-lepton $\ell$. We start with the basic cross section overall normalization results.\par

For reference, we first present in Table~\ref{tab-1} the attendant Born cross sections using the MSTW 2008~\cite{mstw2008} PDF's (we use these PDF's henceforth) based on $10^8$ events.
\begin{table}[h]
\caption{\text{Born Results}}
\begin{center}
\begin{tabular}{|c|c|}
\hline
\text{Before Cuts}&887.797$\pm$ 0.040 \text{pb} \\
\text{With Cuts}& 395.809$\pm$ 0.046 \text{pb}\\
\hline
\end{tabular}
\end{center}
\label{tab-1}
\end{table}
We now move to the comparisons of the results of observables in which the best prediction of the exact ${\cal O}(\alpha^2 L)$ CEEX calculation (labeled CEEX2) is compared with less precise predictions all of which we denote as follows:\\
\begin{itemize}
\item   $ {\cal O}(\alpha^2 L)$ CEEX with ISR+FSR+IFI -- labeled as ``CEEX2''
\item   ${\cal O}(\alpha^2 L)$ CEEX without IFI (initial state final state interference) -- labeled `` CEEX2 (no IFI)''
\item   ${\cal O}(\alpha)$ EEX -- labeled ``EEX1''
\item   ${\cal O}(\alpha)$ EEX without ISR (initial state radiation) -- labeled ``EEX1 (no ISR)''.
\end{itemize}
For further reference, we show in Table~\ref{tab-2} the cross sections with and without the cuts for the four levels of precision which we feature in the studies which follow.
\begin{table}[ht]
\caption{Cross Sections with Higher Order EW Corrections (Matched to a QCD Parton Shower)}
\begin{center}
\begin{tabular}{|c|c|c|}
\hline
~~ & \text{uncut (pb)}& \text{cut (pb)}\\
\hline
\text{CEEX2}& $846.51\pm 0.12$ &$353.69\pm 0.08$\\
\text{CEEX2 (no IFI)}& $846.52\pm 0.12$& $353.63\pm 0.08$\\
\text{EEX1}& $845.87\pm 0.12$& $353.66\pm 0.08$\\
\text{EEX1 (no ISR)}&$845.64\pm 0.05$& $354.94\pm 0.05$\\
%\text{CEEX2}& 844.74 &280.36\\
%\text{CEEX2 (no IFI)}& 844.97& 280.31\\
%\text{EEX1}& 844.45& 280.38\\
%\text{EEX1 (no ISR)}&844.97& 280.64\\
\hline
\end{tabular}
\end{center}
\label{tab-2}
\end{table}
The uncut cases are consistent to 0.077\% whereas the cut cases with (without) ISR are consistent to 0.017\% (0.37\%), respectively. In the uncut cross section,
the we require only that $M_{\ell\ell} > 50$ GeV.\par

\begin{figure}[h]
\begin{center}
\setlength{\unitlength}{1in}
\begin{picture}(6,2.4)(0,0)
\put(0,0.2){\includegraphics[width=3in]{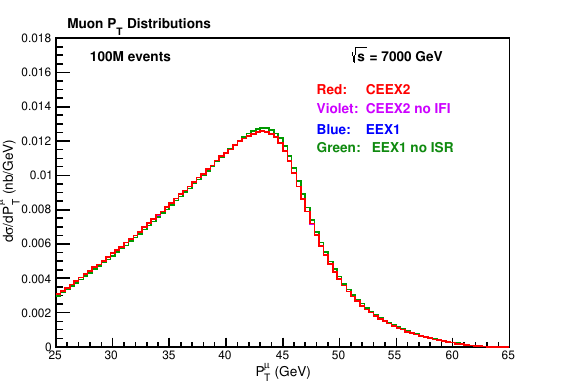}}
\put(3,0.2){\includegraphics[width=3in]{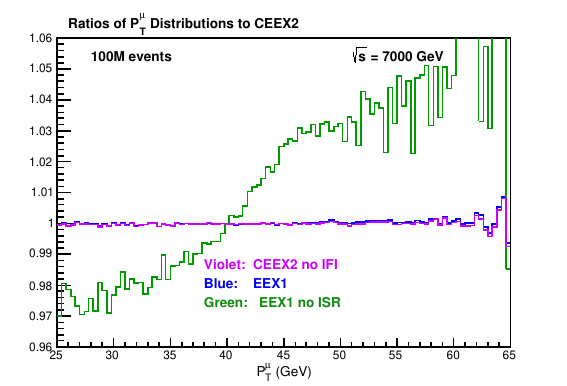}}
\end{picture}
\end{center}
\vspace{-10mm}
\caption{\baselineskip=11pt Muon transverse momentum distributions and their ratios for {\KK}MC-hh with the cuts specified in the text for the EW-CORR (electroweak-correction) labels ``CEEX2" (red -- medium dark shade), ``CEEX2 (no IFI)" (violet -- light dark shade), ``EEX1'' (blue -- dark shade), and ``EEX1 (no IFI)'' (green -- light shade), showered by HERWIG 6.5. The labels are explained in the text. The ratio plot features ``CEEX2'' as the reference distribution as noted in the respective title.}
\label{fig-1}
\end{figure}

We turn next to the muon transverse momentum distribution which we show in Fig.~\ref{fig-1} for $10^8$ events.
In Fig.~\ref{fig-1}, we see that the ISR plays an essential role\footnote{For definiteness and illustration here, we take the quark masses as $m_u=6.0$ MeV,$\; m_d=10.0$ MeV,$\;  m_s= 0.15$ GeV,$\; m_c=1.67$ GeV and $\; m_b=4.78$ GeV, following the analysis in Ref.~\cite{mstw-mass}. In contrast to what is done in Refs.~\cite{wack1,wack2,ditt1,fulv1,ditt2,vicini-wack:2016}, we calculate directly the radiative effects from real photon emission from quarks in the initial state, as these photon quanta are not confined. In Refs.~\cite{wack1,wack2,ditt1,fulv1,ditt2,vicini-wack:2016}, the transverse degrees of freedom of the real photons emitted from the initial state quarks are integrated out so that the big logs L from such emission are absorbed in the quark PDF's in analogy with what is done for gluon emission. In Refs.~\cite{wack1,wack2,ditt1} PDF's with QED evolution were not available for use in the reported phenomenological results. In Refs.~\cite{fulv1,ditt2,vicini-wack:2016}, the PDF's are taken with QED evolution for overall consistency. Since photons are not confined, this approximation that their transverse degrees of freedom may be integrated out can only be trusted for collinear effects in the leading log approximation. The fact that the effect of the QED big logs on the quark and anti-quark PDF's, an example of such a collinear effect, is small does not mean that the effects of the emitted real photons' transverse degrees of freedom are small
on measured observables. The quark PDF's are longitudinal quantities and their changes cannot be used to estimate the effects of the transverse degrees freedom of the radiated photons. In a real sense, the changes in the quark PDF's from QED radiation are red herrings in this discussion.  In our work, we calculate the actual quantum mechanical prediction for the radiation from the initial state quarks without the approximation that the photons' transverse degrees of freedom may be integrated out. To repeat, a detailed comparison with the results in Refs.~\cite{wack1,wack2,ditt1,fulv1,ditt2,vicini-wack:2016} will appear elsewhere.} in modulating the differential lepton momentum spectrum at the few per cent level with a non-flat effect from 25 GeV/c to 65 GeV/c. In this spectrum, generally the effects of the IFI (see the violet (light dark shade) ratio plot) and of the exact ${\cal O}(\alpha^2L)$ correction (see the blue (dark shade) ratio plot) are respectively below and at or below the level of a per mille. Clearly, any truly per mille level study has to take the ISR  into account and, to be safe, such a study should also take the latter two effects into account. 
When we make the last assessment, we do so with the understanding that our results for the latter two effects (the IFI and the  exact ${\cal O}(\alpha^2L)$ correction in Fig. 1)  have uncertainties at the fractional per mille level due to the still imprecise knowledge of the quark masses themselves. As a conservative estimate of the size of the effects due to the uncertainty of the quark masses, we have repeated the calculations in Fig.~\ref{fig-1} in the Appendix as shown in Fig.~\ref{fig-10}
using the PDG~\cite{PDG:2016} values (see the Appendix) for the quark masses. As we explain in the Appendix, while the transverse observable such as the muon $p_T$ is not affected strongly by the attendant change in the quark masses, we do see a non-trivial mass dependence: the ISR still enters at the same level but is shifted at (or below) the 6 per mille level and the sizes of the IFI and the ${\cal O}(\alpha^2L)$ are still at the same level but are shifted at the fractional per mille level\footnote{Here, and henceforward, to quantify the size of the response to the change in the quark masses, we use bins in analogy with Ref.~\cite{atlasmw-17}.}. \par
\par

\begin{figure}[h]
\begin{center}
\setlength{\unitlength}{1in}
\begin{picture}(6,2.4)(0,0)
\put(0,0.2){\includegraphics[width=3in]{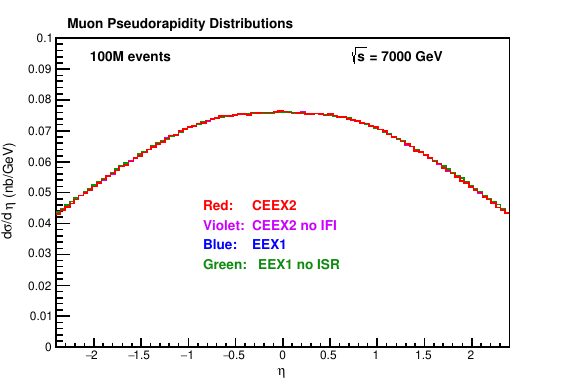}}
\put(3,0.2){\includegraphics[width=3in]{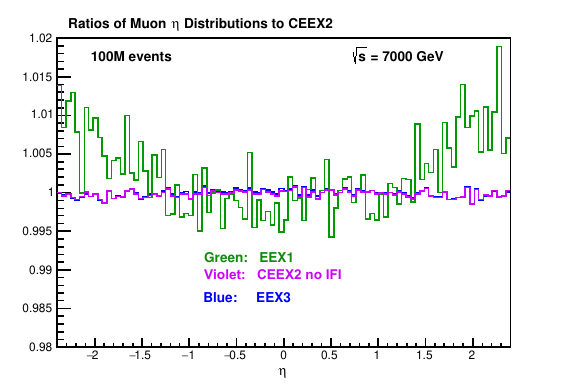}}
\end{picture}
\end{center}
\vspace{-10mm}
\caption{\baselineskip=11pt Muon $\eta$ distributions and their ratios for {\KK}MC-hh with the cuts specified in the text for the EW-CORR labels and notational and illustrative conventions as given in the caption for Fig.~\ref{fig-1}. The events were showered by HERWIG 6.5.The ratio plot features ``CEEX2'' as the reference distribution as noted in the respective title.}
\label{fig-2}
\end{figure}

We consider next the muon $\eta$ distribution as we show it in Fig.~\ref{fig-2} for $10^8$ events.
We see the modulation of the spectrum by the ISR at the 0.5 per cent level while the IFI (see the violet (light dark shade) ratio plot) and the  exact ${\cal O}(\alpha^2L)$ correction (see the blue (dark shade) ratio plot) are at or below the fractional per mille level. Per mille level studies would be advised to take all three effects into account for a conservative precision analysis.
When we repeat the calculations with the PDG~\cite{PDG:2016} values given in the Appendix for the quark masses, we see (in Fig.~\ref{fig-10} in the Appendix) that the respective effects are very similar in size but the ISR effect is shifted at or below the level of a per mille whereas the IFI and the ${\cal O}(\alpha^2L)$ effects are shifted at the fractional per mille level.\par

\begin{figure}[h]
\begin{center}
\setlength{\unitlength}{1in}
\begin{picture}(6,2.4)(0,0)
\put(0,0.2){\includegraphics[width=3in]{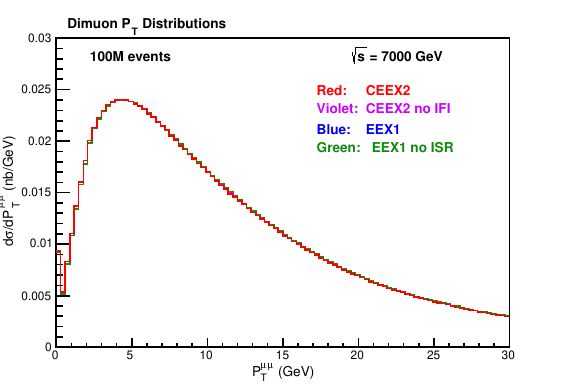}}
\put(3,0.2){\includegraphics[width=3in]{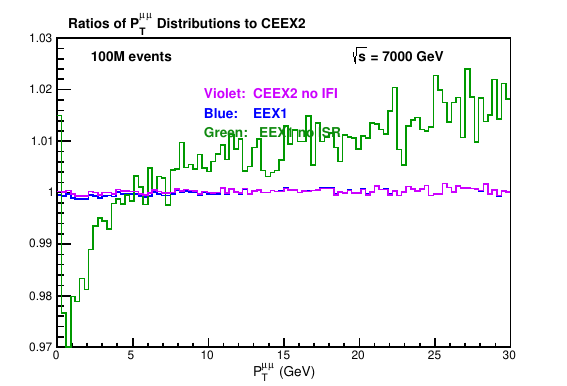}}
\end{picture}
\end{center}
\vspace{-10mm}
\caption{\baselineskip=11pt Dimuon transverse momentum distributions and their ratios for {\KK}MC-hh with the cuts specified in the text for the EW-CORR labels and notational and illustrative conventions as given in the caption for Fig.~\ref{fig-1}. The events were showered by HERWIG 6.5.The ratio plot features ``CEEX2'' as the reference distribution as noted in the respective title.}
\label{fig-3}
\end{figure}

We turn next to the the dimuon transverse momentum distribution which we present in Fig.~\ref{fig-3} for $10^8$ events. 
We see the modulation of the spectrum by the ISR (see the green (light shade) ratio plot) at the per cent level in a non-flat way whereas the IFI and exact ${\cal O}(\alpha^2L)$ effects enter at or below the fraction of a per mille and the per mille level respectively (see the respective violet (light dark shade) and blue (dark shade) ratio plots). Per mille level
studies should take the ISR and, to be conservative, the latter two effects into account in any estimate of overall precision. When we repeat the calculations with the PDG~\cite{PDG:2016} values for the quark masses, we see (in Fig.~\ref{fig-11} in the Appendix) that the respective effects are similar in size and shape but that the ISR effect is shifted at or below the level of 2 per mille while the IFI and exact ${\cal O}(\alpha^2L)$ effects are shifted at the fractional per mille level.\par

\begin{figure}[h]
\begin{center}
\setlength{\unitlength}{1in}
\begin{picture}(6,2.4)(0,0)
\put(0,0.2){\includegraphics[width=3in]{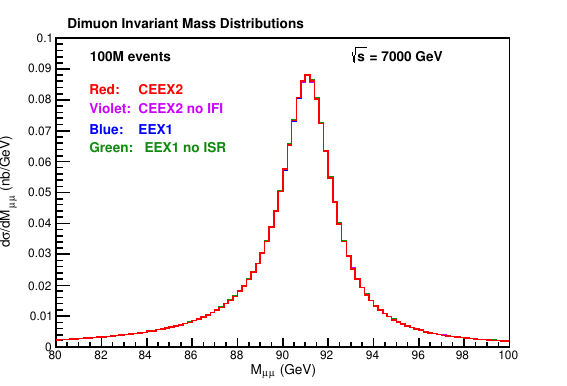}}
\put(3,0.2){\includegraphics[width=3in]{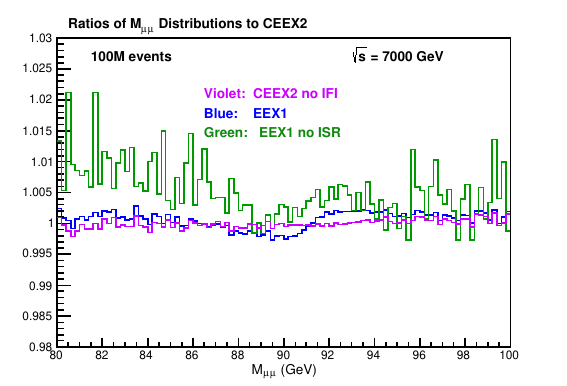}}
\end{picture}
\end{center}
\vspace{-10mm}
\caption{\baselineskip=11pt Dimuon invariant mass distributions and their ratios for {\KK}MC-hh with the cuts specified in the text for the EW-CORR labels and notational and illustrative conventions as given in the caption for Fig.~\ref{fig-1}. The events were showered by HERWIG 6.5.The ratio plot features ``CEEX2'' as the reference distribution as noted in the respective title.}
\label{fig-4}
\end{figure}

We consider next the dimuon invariant mass spectrum which we present in Fig.~\ref{fig-4} for $10^8$ events.
In this spectrum, the ISR modulation (see the green (light shade) ratio plot) exceeds 1\% at the lower mass values and is non-flat in shape whereas the IFI (see the violet (light dark shade) ratio plot) reaches the per mille level, in a non-flat shape, at the higher mass values and the exact ${\cal O}(\alpha^2L)$ correction (see the blue (dark shade) ratio plot) enters at the few per mille level in a non-flat shape.
Per mille level studies need to take the ISR and the exact ${\cal O}(\alpha^2L)$ correction into account and, to be conservative, need to take the IFI into account. 
When we repeat the calculations with the PDG~\cite{PDG:2016} values for the quark masses
we see (in Fig.~\ref{fig-11} in the Appendix) the similar size effects with a somewhat stronger effect for the ISR by a few per mille.
\par
\begin{figure}[h]
\begin{center}
\setlength{\unitlength}{1in}
\begin{picture}(6,2.4)(0,0)
\put(0,0.2){\includegraphics[width=3in]{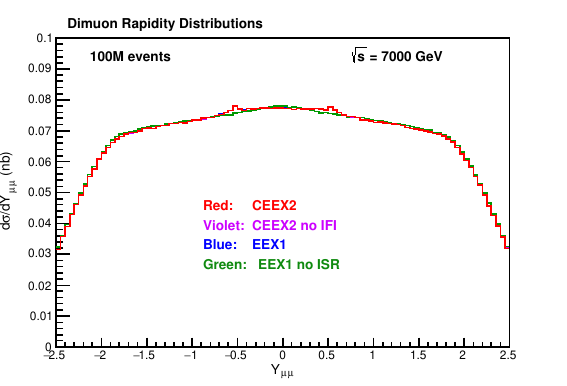}}
\put(3,0.2){\includegraphics[width=3in]{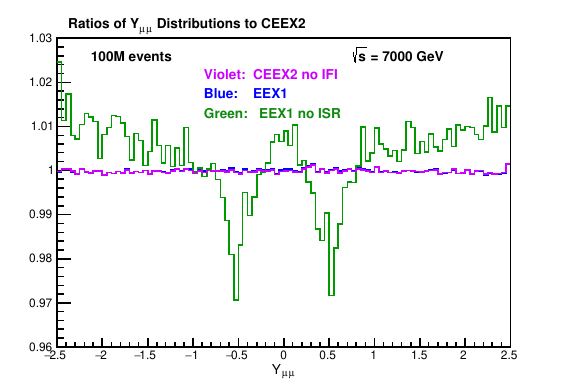}}
\end{picture}
\end{center}
\vspace{-10mm}
\caption{\baselineskip=11pt Dimuon rapidity distributions and their ratios for {\KK}MC-hh with the cuts specified in the text for the EW-CORR labels and notational and illustrative conventions as given in the caption for Fig.~\ref{fig-1}.The ratio plot features ``CEEX2'' as the reference distribution as noted in the respective title.}
\label{fig-5}
\end{figure}

We show in Fig.~\ref{fig-5} the dimuon rapidity distribution for $10^8$ events, where the events are showered with HERWIG 6.5.
For the ISR (see the green (light shade) ratio plot) we see non-flat modulations at the per cent level while for the IFI (see the violet (light dark shade) ratio plot) and for the exact ${\cal O}(\alpha^2L)$ correction (see the blue (dark shade) ratio plot) we have at most fractional per mille level modulations. The ISR should definitely be taken into account by per mille level studies. A more conservative strategy would to take all three effects account in per mille level precision estimates. When we repeat the calculations (see the Appendix, Fig.~\ref{fig-12}) for the PDG~\cite{PDG:2016} quark mass values we see similar size effects, with a per mille level enhancement of the ISR effect. \par

\begin{figure}[h]
\begin{center}
\setlength{\unitlength}{1in}
\begin{picture}(6,2.4)(0,0)
\put(0,0.2){\includegraphics[width=3in]{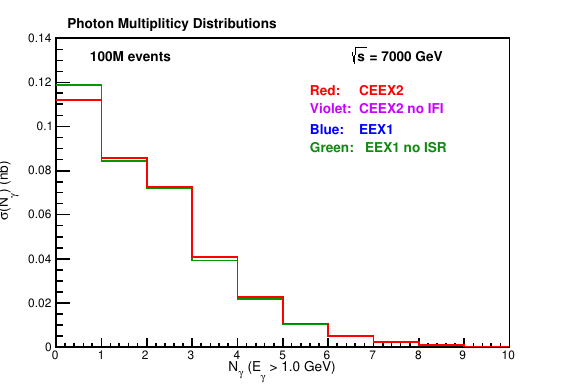}}
\put(3,0.2){\includegraphics[width=3in]{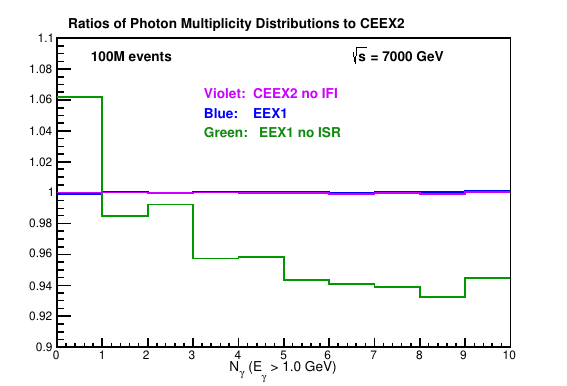}}
\end{picture}
\end{center}
\vspace{-10mm}
\caption{\baselineskip=11pt Photon multiplicity distributions and their ratios for photons with energy $>$ 1 GeV for {\KK}MC-hh with the cuts specified in the text for the EW-CORR labels and notational and illustrative conventions as given in the caption for Fig.~\ref{fig-1}.The ratio plot features ``CEEX2'' as the reference distribution as noted in the respective title.}
\label{fig-6}
\end{figure}

We turn next in Fig.~\ref{fig-6} to the total photon multiplicity distribution for photons with energy $>$ 1 GeV, for $10^8$ events which were showered by HERWIG 6.5.
For the ISR (see the green (light shade) ratio plot) there is non-flat modulation at the level of 5 per cent while the IFI (see the violet (light dark shade) ratio plot) and the exact ${\cal O}(\alpha^2L)$ correction (see the blue (dark shade) ratio plot) are generally within fractional per mille of the reference ``CEEX2''. Per mille level studies should definitely take the ISR into account.
When we repeat the calculations (see the Appendix, Fig.~\ref{fig-13}) for the PDG~\cite{PDG:2016} quark mass values we see similar size effects, with a 5 per mille level enhancement of the ISR effect.
\par

\begin{figure}[h]
\begin{center}
\setlength{\unitlength}{1in}
\begin{picture}(6,2.4)(0,0)
\put(0,0.2){\includegraphics[width=3in]{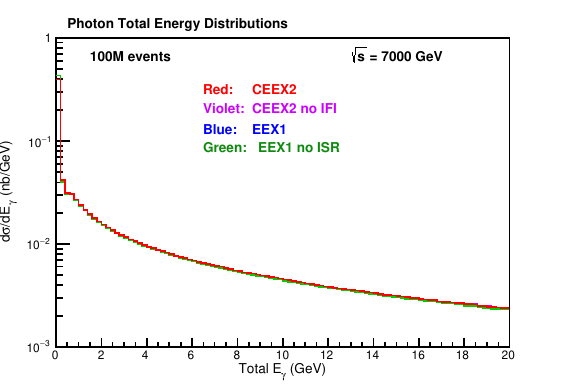}}
\put(3,0.2){\includegraphics[width=3in]{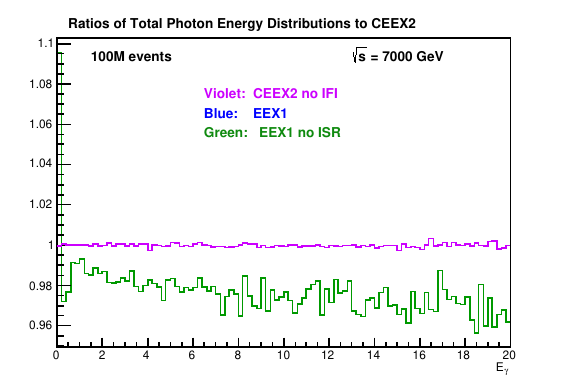}}
\end{picture}
\end{center}
\vspace{-10mm}
\caption{\baselineskip=11pt Total photon energy distributions and their ratios for photons with energy $>$ 1 GeV for {\KK}MC-hh with the cuts specified in the text for the EW-CORR labels and notational and illustrative conventions as given in the caption for Fig.~\ref{fig-1}.The ratio plot features ``CEEX2'' as the reference distribution as noted in the respective title.}
\label{fig-7}
\end{figure}

In Fig.~\ref{fig-7} we show the distributions for the total photon energy for $10^8$ events showered by HERWIG 6.5.
For the ISR (see the green (light shade) ratio plot) we have a non-flat modulation at the few per cent level while for the IFI (see the violet (light dark shade) ratio plot) we have a non-flat modulation at or below the fractional per mille level and for the exact ${\cal O}(\alpha^2L)$ correction (see the blue (dark shade) ratio plot) we have a similar non-flat fractional per mille level modulation. Per mille level studies should take the ISR and the  exact ${\cal O}(\alpha^2L)$ correction into account. A more conservative approach would take all three effects should be taken into account in per mille level studies. When we repeat the calculations (see the Appendix, Fig.~\ref{fig-13}) for the PDG~\cite{PDG:2016} quark mass values we see similar size effects, with a 2 per mille level modulation of the ISR effect in the regime of 2 GeV.
\par

\begin{figure}[h]
\begin{center}
\setlength{\unitlength}{1in}
\begin{picture}(6,2.4)(0,0)
\put(0,0.2){\includegraphics[width=3in]{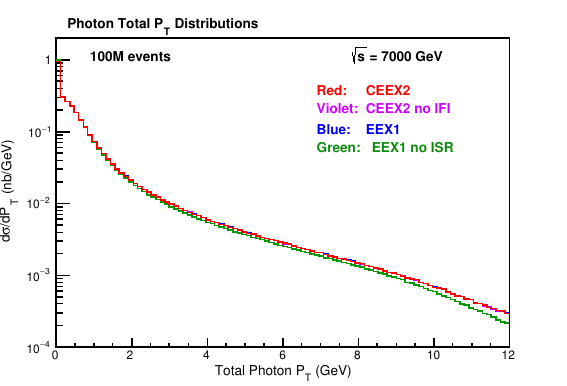}}
\put(3,0.2){\includegraphics[width=3in]{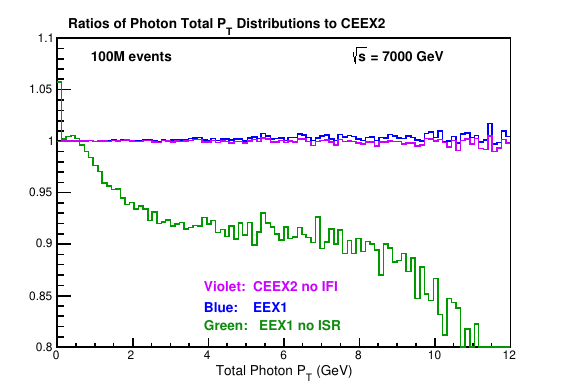}}
\end{picture}
\end{center}
\vspace{-10mm}
\caption{\baselineskip=11pt Total photon transverse momentum distributions and their ratios for photons with energy $>$ 1 GeV for {\KK}MC-hh with the cuts specified in the text for the EW-CORR labels and notational and illustrative conventions as given in the caption for Fig.~\ref{fig-1}.The ratio plot features ``CEEX2'' as the reference distribution as noted in the respective title.}
\label{fig-8}
\end{figure}

In Fig.~\ref{fig-8} we consider the total transverse momentum distribution of photons for $10^8$ events which were showered by HERWIG 6.5.
For the ISR (see the green (light shade) ratio plot) we have non-flat effects at that reach the 15 per cent level whereas for the IFI (see the violet (light dark shade) ratio plot) the effects are non-flat and at or below the per mille level. For the exact ${\cal O}(\alpha^2L)$ correction (see the blue (dark shade) ratio plot) the effects are at the few per mille level and are non-flat. Per mille level studies, to be conservative, should take all three effects into account.  When we repeat the calculations (see the Appendix, Fig.~\ref{fig-12}) for the PDG~\cite{PDG:2016} quark mass values we see similar size effects with the entirely similar characters wherein the ISR effect is shifted at the level of 1\% while the IFI and exact ${\cal O}(\alpha^2L)$ effects are shifted by fractional per mille levels.
\par

\begin{figure}[h]
\begin{center}
\setlength{\unitlength}{1in}
\begin{picture}(6,2.4)(0,0)
\put(0,0.2){\includegraphics[width=3in]{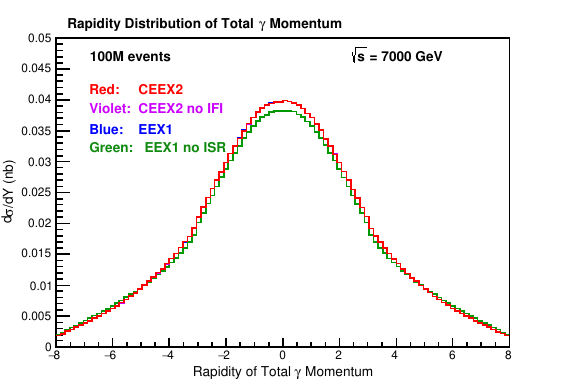}}
\put(3,0.2){\includegraphics[width=3in]{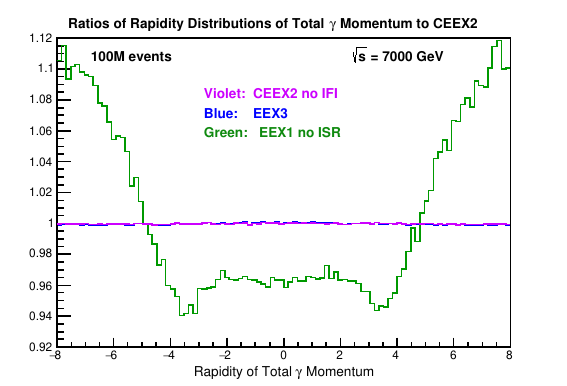}}
\end{picture}
\end{center}
\vspace{-10mm}
\caption{\baselineskip=11pt Photon total momentum rapidity distributions and their ratios for photons with energy $>$ 1 GeV for {\KK}MC-hh with the cuts specified in the text for the EW-CORR labels and notational and illustrative conventions as given in the caption for Fig.~\ref{fig-1}.The ratio plot features ``CEEX2'' as the reference distribution as noted in the respective title.}
\label{fig-9}
\end{figure}

In Fig.~\ref{fig-9} we turn to the rapidity of the total photon momentum for $10^8$ events showered by HERWIG 6.5.
The ISR effect (see the green (light shade) ratio plot) is non-flat and at the level of 4 per cent in the central region and at the level of 8 per cent in the forward/backward regions whereas the IFI effect (see the violet (light dark shade) ratio plot) is at or below the level of fractional per mille and the exact ${\cal O}(\alpha^2L)$ correction (see the blue (dark shade) ratio plot) is at or below the level of a per mille and is non-flat. Precision studies at the per mille level should take the ISR and the  exact ${\cal O}(\alpha^2L)$ correction into account. To be more conservative, per mille level studies should take all three effects into account. When we repeat the calculations (see the Appendix, Fig.~\ref{fig-14}) for the PDG~\cite{PDG:2016} quark mass values we see similar size effects wherein the ISR effect is shifted by the level of 6 per mille (1 per mille) in the forward/backward (central) regions and the IFI and exact ${\cal O}(\alpha^2L)$ effects are shifted by fractional per mille levels.
%\footnote{That our results are rather robust to the precise values used for the quark masses (The PDG values for the light quarks differ by a factor of $\sim 2$ from the values we use in the text.) is important. This is discussed quantitatively in the Appendix. Ultimately, we expect that the lattice-based methods will reduce the error on the quark masses we use here to the level of a few per cent~\cite{pmac,nplat}.}
\par
%%%STARTHERE
%\section{Comparisons with the Results in Ref.~\cite{vicini-wack}}

%In this section we make contact with the study in Ref.~\cite{vicini-wack} in which the results for several available exact results and MC event generators with higher order EW corrections were cataloged in a set of tuned comparisons. In making contact with the results in Ref.~\cite{vicini-wack} we focus on the effects beyond the exact NLO EW corrections since the
%exact NLO EW corrections in {\KK}MC-hh are based on the same DIZET library that was already featured in the SANC~\cite{sanc} results in the study in Ref.~\cite{vicini-wack}.\par

%To be sure that we have the benchmark setups properly implemented in {\KK}MC-hh, we start by checking that we get the same Born cross sections as found in the respective part of the study in Ref.~\cite{vicini-wack}. 

\section{Summary}
 
What we have shown in our discussion here, using as an illustrator the $Z/\gamma^*$ spectra used in the pioneering analysis in Ref.~\cite{atlasmw-17}, is the need to take various higher order EW effects, as illustrated using the {\KK}MC-hh, into account in precision studies of heavy $Z/\gamma^*$ with decay to lepton pairs at the LHC. Specifically, the ISR is the most pronounced
effect, where it can be as large as several per cent in some observables. The exact  ${\cal O}(\alpha^2 L)$ corrections can reach several per mille in some observables and the IFI is generally at or below the fractional per mille level. When we repeat, in the Appendix, our calculations using the PDG~\cite{PDG:2016} quark mass values (The PDG values for the light quarks differ by a factor of $\sim 2$ from the values we use in the main text.) we see similar size effects but with shifts at the level of $\sim 10\%$ of the size of the effects shown in the main text, in accordance with the size of the change in the respective big log $L$. Ultimately, we expect that the lattice-based methods will reduce the error on the quark masses we use here to the level of a few per cent~\cite{pmac,nplat}. When the precision tag is at the per mille level, the ISR, IFI and exact ${\cal O}(\alpha^2 L)$ corrections should be
included in the analysis for a conservative treatment of the respective precision estimate.Toward this end, the  {\KK}MC-hh MC is available from the authors upon request.\par
 
\vskip 2 mm
\centerline{\bf Acknowledgments}
\vskip 2 mm

This work was supported in part by the Polish National Centre of Science 
Grants No. DEC-2011/03/B/ST2/00220 and DEC-2012/04/M/ST2/00240,
by the Programme of the French–Polish Cooperation between IN2P3 and COPIN 
within the Collaborations Nos. 10-138 and 11-142 and by a grant from the Citadel Foundation. The authors also thank Prof. G. Giudice for the support and kind hospitality of the CERN TH Department. 
\section*{Appendix}
 In this Appendix we record the results which would obtain in Figs. ~1-9 if one uses the 
\begin{figure}[h]
\begin{center}
\setlength{\unitlength}{1in}
\begin{picture}(6,3.4)(0,0)
\put(0,0.2){\includegraphics[width=6in]{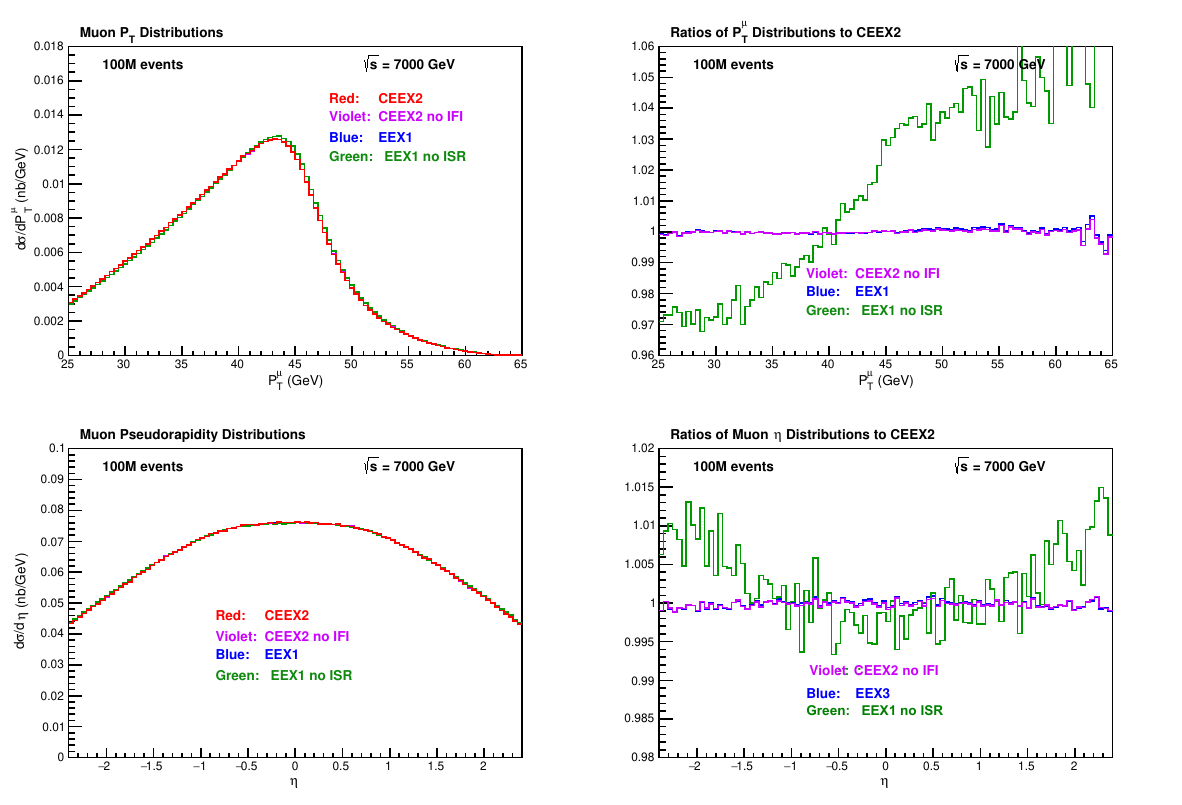}}
%\put(0,0.2){\includegraphics[width=3in]{photon-rapidity-6-18-18.pdf}}
%\put(3,0.2){\includegraphics[width=3in]{photon-rapidity-ratio-6-18-18.pdf}}
\end{picture}
\end{center}
\vspace{-10mm}
\caption{\baselineskip=8pt Muon transverse momentum and pseudorapidity distributions and their respective ratios, using the PDG quark masses as explained in the text, for photons with energy $>$ 1 GeV for {\KK}MC-hh with the cuts specified in the text for the EW-CORR labels and notational and illustrative conventions as given in the caption for Fig.~\ref{fig-1}.The ratio plot features ``CEEX2'' as the reference distribution as noted in the respective title.}
\label{fig-10}
\end{figure}
PDG~\cite{PDG:2016} quark masses $(m_u = 2.2^{+0.6}_{-04.}\text{MeV},\; m_d = 4.7^{+0.5}_{-0.4}\text{MeV},\; m_s = 96^{+8}_{-4}\text{MeV})$ instead of those used in the text.
\begin{figure}[h!]
\begin{center}
\setlength{\unitlength}{1in}
\begin{picture}(6,3.4)(0,0)
\put(0,0.2){\includegraphics[width=6in]{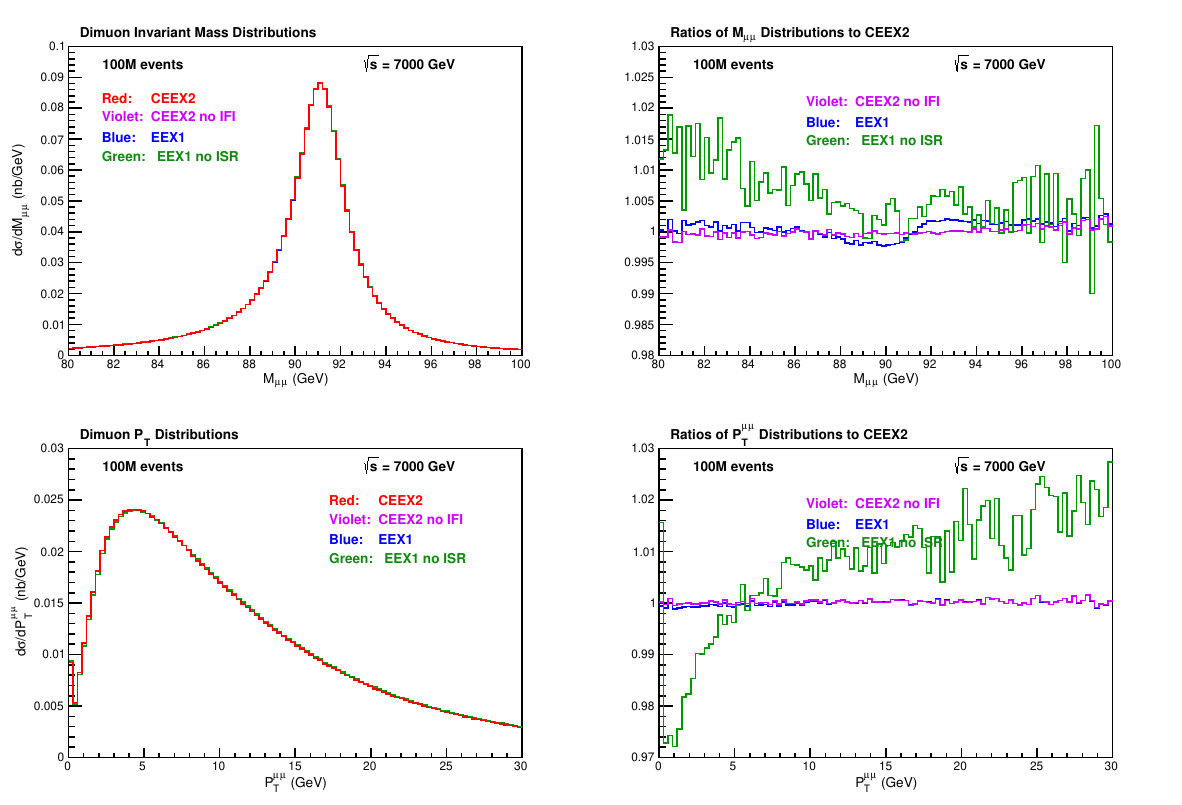}}
%\put(0,0.2){\includegraphics[width=3in]{photon-rapidity-6-18-18.pdf}}
%\put(3,0.2){\includegraphics[width=3in]{photon-rapidity-ratio-6-18-18.pdf}}
\end{picture}
\end{center}
%\vspace{-10mm}
\caption{\baselineskip=8pt Dimuon mass and transverse momentum distributions and their respective ratios, using the PDG quark masses as explained in the text, for {\KK}MC-hh with the cuts specified in the text for the EW-CORR labels and notational and illustrative conventions as given in the caption for Fig.~\ref{fig-1}.The ratio plot features ``CEEX2'' as the reference distribution as noted in the respective title.}
\label{fig-11}
\end{figure}
We see in Figs.~10-14 that the size of the effects discussed in the text are not substantially affected,
\begin{figure}[t]
\begin{center}
\setlength{\unitlength}{1in}
\begin{picture}(6,3.4)(0,0)
\put(0,0.2){\includegraphics[width=6in]{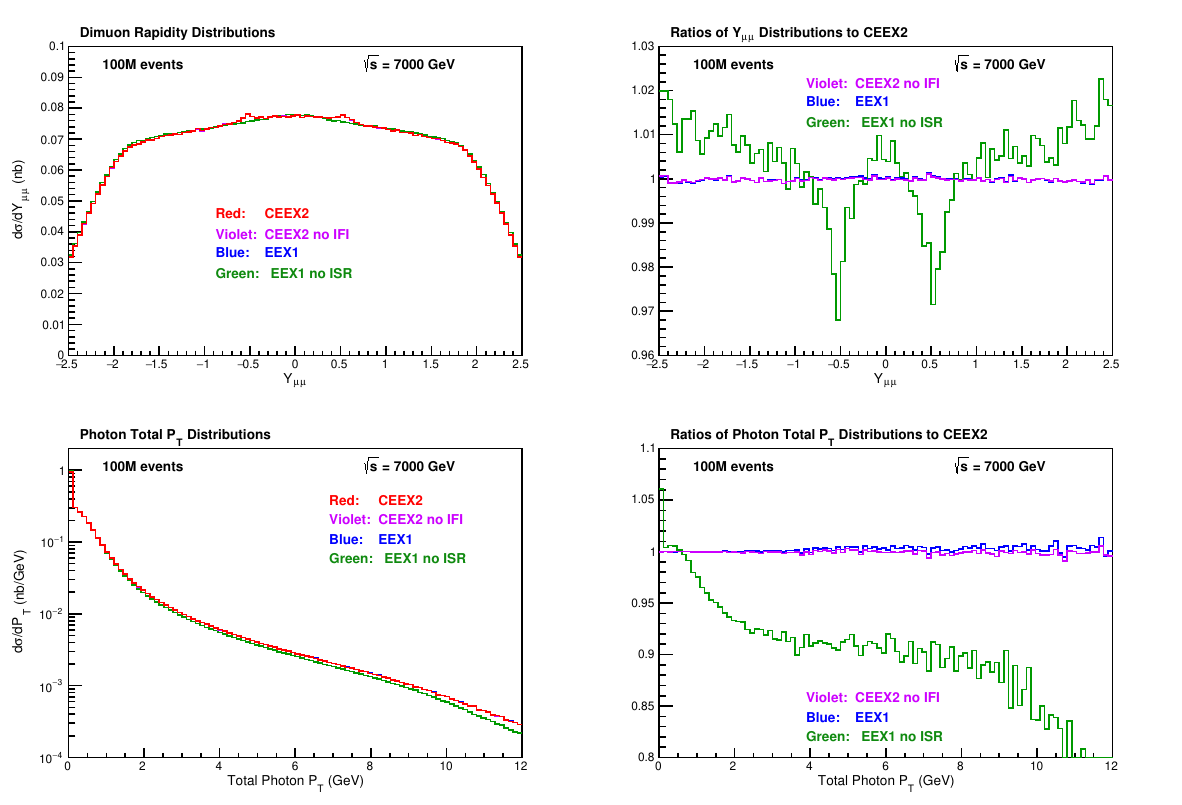}}
%\put(0,0.2){\includegraphics[width=3in]{photon-rapidity-6-18-18.pdf}}
%\put(3,0.2){\includegraphics[width=3in]{photon-rapidity-ratio-6-18-18.pdf}}
\end{picture}
\end{center}
\vspace{-10mm}
\caption{\baselineskip=8pt Dimuon rapidity and photon total transverse momentum distributions and their respective ratios, using the PDG quark masses as explained in the text, for photons with energy $>$ 1 GeV for {\KK}MC-hh with the cuts specified in the text for the EW-CORR labels and notational and illustrative conventions as given in the caption for Fig.~\ref{fig-1}.The ratio plot features ``CEEX2'' as the reference distribution as noted in the respective title.}
\label{fig-12}
\end{figure}
 especially when one recalls that the PDG values correspond to a scale of 2 GeV\footnote{To estimate the size of the uncertainty due to the uncertainty of the current quark masses, we compare the size of the logarithmic ISR effect for two different values of the masses, the value in Ref.~\cite{mstw-mass} and the PDG values. If we do this, we get an effect of the size  $(2\frac{\alpha}{\pi})Q_u^2 \ln(m_1/m_2) = 2\frac{\alpha}{\pi}(4/9)\ln(6/2.2) = 0.21\%$ for the u quark case and of the size $0.039\%$ for the d-quark case. Since we have roughly the same number of u and d quark events in our simulation, this finally averages to $0.12\%$.}.
\begin{figure}[t]
\begin{center}
\setlength{\unitlength}{1in}
\begin{picture}(6,3.4)(0,0)
\put(0,0.2){\includegraphics[width=6in]{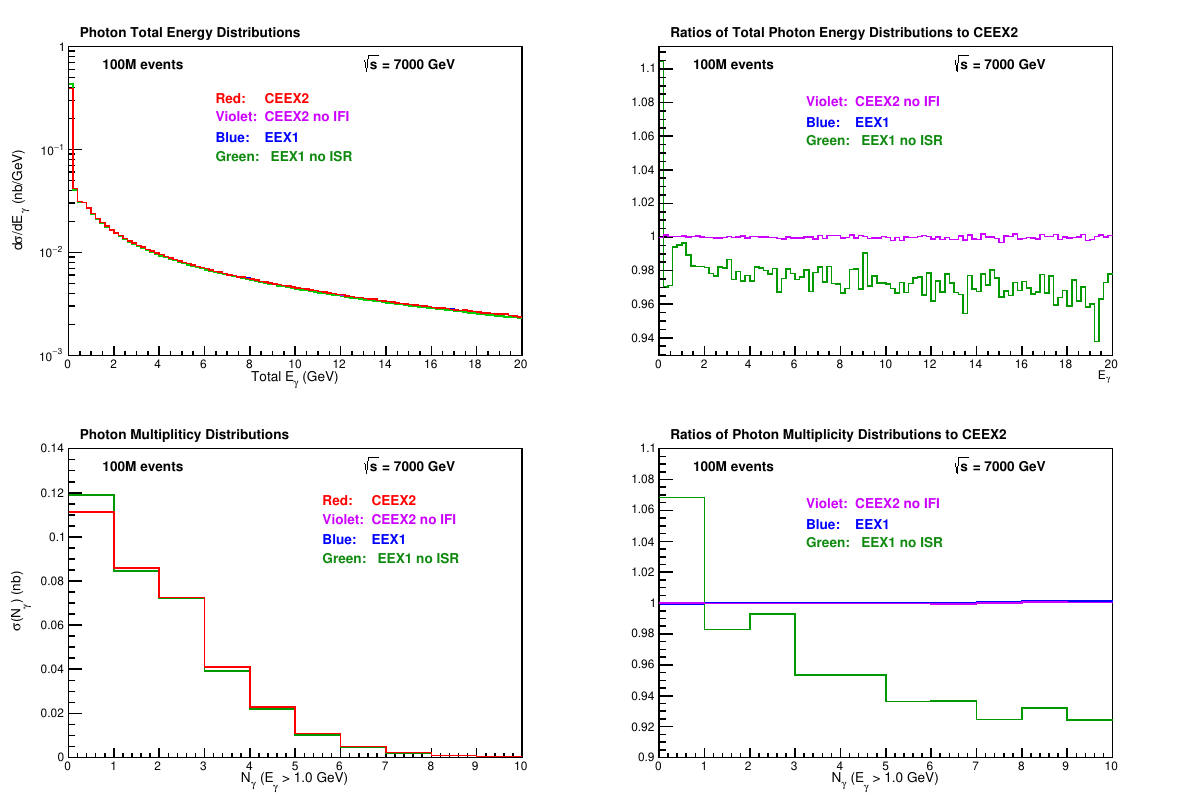}}
%\put(0,0.2){\includegraphics[width=3in]{photon-rapidity-6-18-18.pdf}}
%\put(3,0.2){\includegraphics[width=3in]{photon-rapidity-ratio-6-18-18.pdf}}
\end{picture}
\end{center}
%\vspace{-10mm}
\caption{\baselineskip=8pt Photon total energy and multiplicity distributions and their respective ratios, using the PDG quark masses as explained in the text, for photons with energy $>$ 1 GeV for {\KK}MC-hh with the cuts specified in the text for the EW-CORR labels and notational and illustrative conventions as given in the caption for Fig.~\ref{fig-1}. The ratio plot features ``CEEX2'' as the reference distribution as noted in the respective title.}
\label{fig-13}
\end{figure}

\begin{figure}[t]
\begin{center}
\setlength{\unitlength}{1in}
\begin{picture}(6,3.4)(0,0)
\put(-.2,-2.0){\includegraphics[width=6.5in]{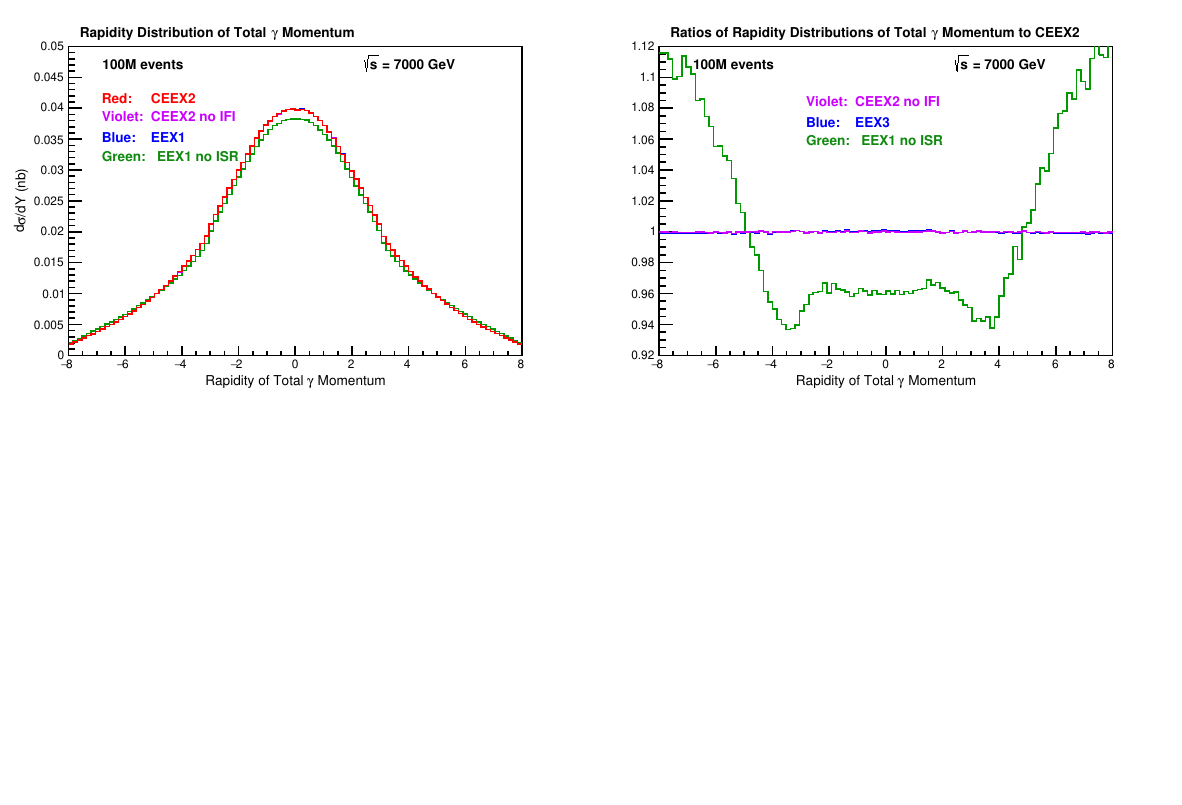}}
%\put(0,0.2){\includegraphics[width=3in]{photon-rapidity-6-18-18.pdf}}
%\put(3,0.2){\includegraphics[width=3in]{photon-rapidity-ratio-6-18-18.pdf}}
\end{picture}
\end{center}
\vspace{-10mm}
\caption{\baselineskip=8pt Photon total momentum rapidity distributions and their respective ratios, using the PDG quark masses as explained in the text, for photons with energy $>$ 1 GeV for {\KK}MC-hh with the cuts specified in the text for the EW-CORR labels and notational and illustrative conventions as given in the caption for Fig.~\ref{fig-1}. The ratio plot features ``CEEX2'' as the reference distribution as noted in the respective title.}
\label{fig-14}
\end{figure} 
\par
\bibliography{Tauola_interface_design}{}
%\bibliography{BU-HEPP-17-01}{}
\bibliographystyle{utphys_spires}

%\end{thebibliography}
%\section{}
%\label{}

%% The Appendices part is started with the command \appendix;
%% appendix sections are then done as normal sections
%% \appendix

%% \section{}
%% \label{}

%% References
%%
%% Following citation commands can be used in the body text:
%% Usage of \cite is as follows:
%%   \cite{key}         ==>>  [#]
%%   \cite[chap. 2]{key} ==>> [#, chap. 2]
%%

%% References with BibTeX database:
%\nocite{*}
%\bibliographystyle{elsarticle-num}
%\bibliography{martin}

%% Authors are advised to use a BibTeX database file for their reference list.
%% The provided style file elsarticle-num.bst formats references in the required Procedia style

%% For references without a BibTeX database:

% \begin{thebibliography}{00}

%% \bibitem must have the following form:
%%   \bibitem{key}...
%%

% \bibitem{}

% \end{thebibliography}

\end{document}